# Giant thermoelectric effect in Al$_2$O$_3$ magnetic tunnel junctions

**Weiwei Lin (林维维)[†], Michel Hehn, Laurent Chaput, Béatrice Negulescu[‡], Stéphane Andrieu, François Montaigne and Stéphane Mangin***

**Thermoelectric effects in magnetic nanostructures and the so-called spin caloritronics are attracting much interests.[1–11,27,28] Indeed it provides a new way to control and manipulate spin currents which are key elements of spin-based electronics.[12,13] Here we report on the giant magnetothermoelectric effect in a magnetic tunnel junction. The thermovoltage in this geometry can reach 1 mV. Moreover a magneto-thermovoltage effect could be measured with ratio similar to the tunnel magnetoresistance ratio. The Seebeck coefficient can then be tuned by changing the relative magnetization orientation of the two magnetic layers in the tunnel junction. Therefore our experiments extend the range of spintronic devices application to thermoelectricity and provide a crucial piece of information for understanding the physics of thermal spin transport.**

Thermoelectricity has been known since 1821 with T.J. Seebeck. On one hand, the relation between the thermal and the electrical transport is an essential topic for both fundamental physics and for the future of energy-saving technologies.[14,15] On the other hand the discovery of the giant magnetoresistance effect (GMR) and the tunnel magnetoresistance effect (TMR) enhanced the interest of the community for spin-dependent conductivity and gave rise to spintronics and multiple applications.[12,13] Its interplay with thermal conductivity was introduced to describe the conventional Seebeck effect in ferromagnetic metals.[1–9,16–20]. The magnetothermoelectric effect has then be studied in magnetic systems such as magnetic multilayers and spin valves.[16–20] Moreover the thermoelectric effect has also been observed in non magnetic tunneling devices such as superconductor-insulator-normal metal (or superconductor) tunnel junctions.[21,22] Recently, thermal spin tunnelling effect from ferromagnet to silicon has been reported.[23] Concerning magnetic tunnel junction (MTJ), there were theoretical works[24-26] showing magnetothermopower, and Walter *et al.*[27] reported firstly the measurements of Seebeck effect in MgO MTJs. Their experiments show that the magnitude and sign of the magneto-Seebeck ratio can be changed by laser power modulation[27].

In this article, we present an experimental discovery of the giant thermoelectric effect in Al$_2$O$_3$ MTJs. The observed mV thermovoltage has promising application for the novel magnetic thermoelectric devices.

The studied MTJ consists of a bottom reference layer Ta(5 nm)/PtMn(25 nm)/Co$_{90}$Fe$_{10}$(2 nm)/Ru(0.8 nm)/Co$_{90}$Fe$_{10}$(3 nm) and a free layer Co$_{90}$Fe$_{10}$(2 nm)/Ni$_{80}$Fe$_{20}$(5 nm)/Ru(4.8 nm)/Au(10 nm) separated by a 2 nm thick amorphous Al$_2$O$_3$ barrier, as shown in Fig. 1a. To generate a temperature difference between the reference layer and the free layer, one electrode lead was heated using the laser beam from a laser diode with a wavelength of 780 nm and a tunable power from 0 to 125 mW. The temperature difference between both sides of the Al$_2$O$_3$ barrier is defined as $\Delta T$ whereas the voltage difference is $\Delta V$. In the linear response approximation, the total electric current $I$ in the presence of $\Delta V$ and $\Delta T$ can be written as[7,16]

$$I = G_V \Delta V + G_T \Delta T \qquad (1)$$

where $G_V$ is the electrical conductance, and $G_T$ is the thermoelectric coefficient related to the charge current response to the heat flux.

The thermovoltage $\Delta V$ can be measured in an open-circuit geometry where $I = 0$, as shown in Fig. 1b. Considering equation (1) it leads to $\Delta V = -(G_T/G_V) \Delta T = -S \Delta T$, where $S = G_T/G_V$ is the thermopower (TP) or Seebeck coefficient. $\Delta V$ was measured with a nanovoltmeter at room temperature (RT) with a magnetic field $H$ applied along the in-plane easy axis of the free layer. The thermotunnel current was measured by a source-meter connecting the MTJ without any applied voltage, i.e. a closed-circuit, as shown in Fig. 1c. In the closed-circuit geometry, $\Delta V = 0$ and thus from equation (1), $I = G_T \Delta T$. With those two geometries the influence of magnetization orientations on both spin-dependent electrical conductivity and thermoelectric effect could be studied.

Figure 2a shows a minor loop of the tunnel resistance $R$ as a function of the in-plane applied field $H$ for an Al$_2$O$_3$ based MTJ with a diameter of 80 µm. The MTJ has a low resistance $R_P = 15.9$ kΩ for the parallel (P) magnetizations alignment, and a high resistance $R_{AP} = 22.3$ kΩ for the antiparallel (AP) magnetizations alignment, showing a TMR ratio $(R_{AP} - R_P)/R_P = 40\%$.

Then, instead of injecting a current in the MTJ, as sketched in Fig. 1b, the voltage across the MTJ is measured in an open-circuit geometry. The top lead is heated by the laser in order to generate a temperature difference between the free layer and the reference layer spaced by the Al$_2$O$_3$ barrier. With the top lead heated, the temperature difference is defined as positive $\Delta T > 0$. As

---

Institut Jean Lamour, Nancy-Université, Vandoeuvre-lès-Nancy 54506, France. [†]Present address: Institut d'Electronique Fondamentale, Université Paris-Sud, Orsay 91405, France. E-mail: weiwei.lin@u-psud.fr ; [‡]Present address: Laboratoire d'Electrodynamique des Matériaux Avancés, CNRS-CEA, Tours 37200, France; *e-mail: stephane.mangin@ijl.nancy-universite.fr.

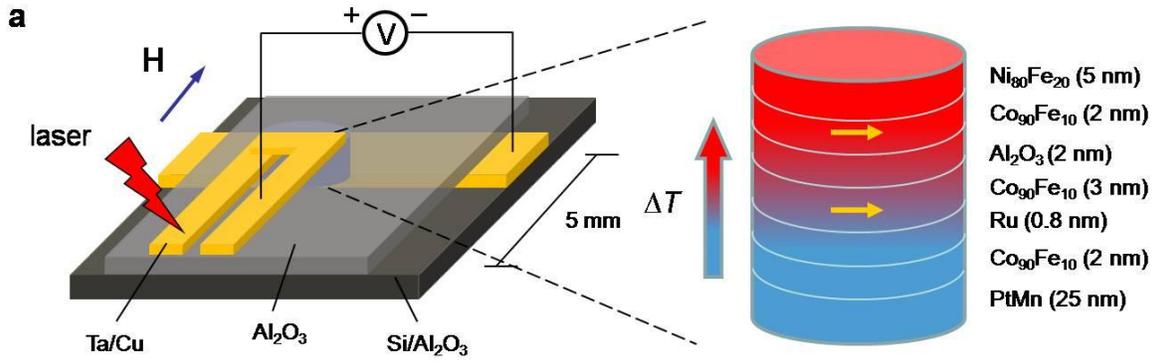

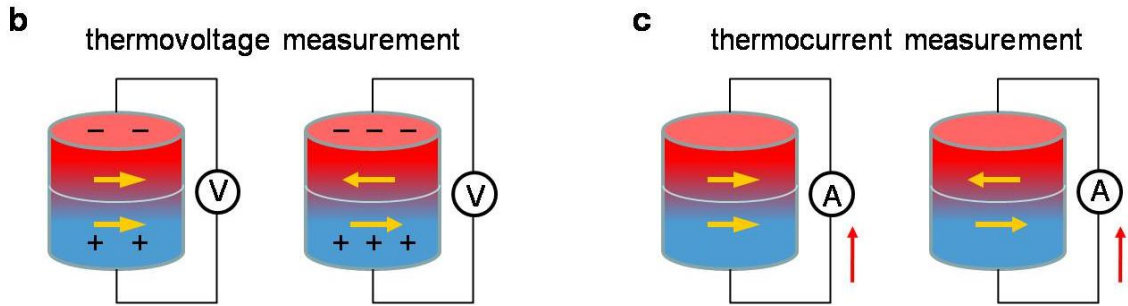

**Figure 1 Schematic of the experiment. a**, The studied MTJ consists of a bottom reference layer Ta(5 nm)/PtMn(25 nm)/Co$_{90}$Fe$_{10}$(2 nm)/Ru(0.8 nm)/Co$_{90}$Fe$_{10}$(3 nm) and a free layer Co$_{90}$Fe$_{10}$(2 nm)/Ni$_{80}$Fe$_{20}$(5 nm)/Ru(4.8 nm)/Au(10 nm) separated by an Al$_2$O$_3$ barrier. To generate a temperature difference between both sides of the Al$_2$O$_3$ barrier, one electrode lead was heated using the laser beam from a laser diode with the wavelength of 780 nm and a maximum power of 125 mW. The open-circuit voltage was measured by the nanovoltmeter at room temperature (RT) with an applied magnetic field $\mu_0 H$ up to 0.3 T along the in-plane easy axis. **b**, In the presence of the temperature difference $\Delta T$ in the MTJ, the generated thermovoltage $\Delta V$ depends on the relative magnetization alignment of the two ferromagnetic layers. **c**, The thermotunnel current $I$ was measured by the system sourcemeter connecting the MTJ in a closed-circuit without applied voltage.

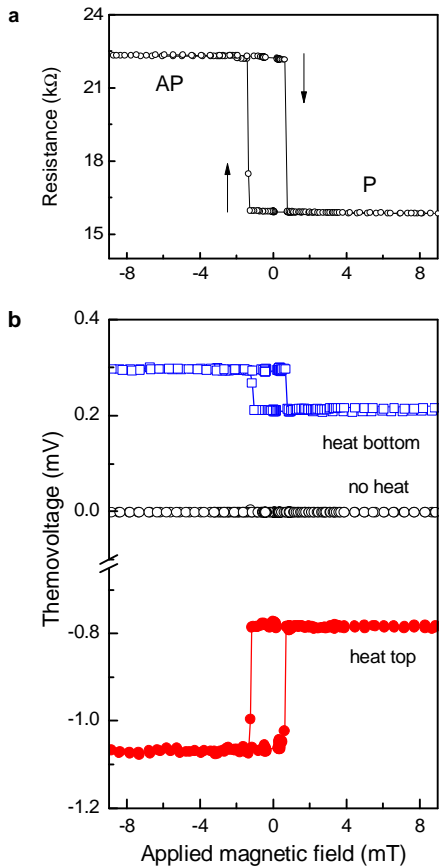

**Figure 2 Magnetic field dependence of the tunnel resistance and the thermovoltage in a MTJ. a**, Minor loop of the tunnel resistance $R$ of an Al$_2$O$_3$ MTJ with a diameter of 80 μm as a function of $H$ at room temperature, measured with a 0.1 μA current. The MTJ has a low resistance $R_P$ = 15.9 kΩ for the parallel (P) magnetization alignment, and a high resistance $R_{AP}$ = 22.3 kΩ for the antiparallel (AP) alignment. The TMR ratio ($R_{AP} - R_P)/R_P$ = 40%. **b**, Thermovoltage vs. applied field ($V$-$H$) minor loops. The voltage across the MTJ is measured in an open-circuit geometry with a laser heating the electrodes. As the laser heats the top lead, the temperature of the free layer is higher than that of the reference one, i.e. $\Delta T > 0$, yielding a negative thermovoltage $\Delta V$, whereas a positive $\Delta V$ is observed in the case of the laser heating the bottom lead, i.e. $\Delta T < 0$. It is noted that the open-circuit voltage is zero in the absence of laser heating. With sweeping the applied field, $\Delta V$ shows a behavior similar to $R$. The amplitude of the thermovoltage for the AP alignment, $\Delta V_{AP}$, is larger than that for the P one, $\Delta V_P$. The $\Delta V_{AP}$ can reach − 1.07 mV as heating the top lead with a 125 mW laser power, while the $\Delta V_{AP}$ is about 310 μV as heating the bottom lead with the same laser power. The tunnel magnetothermovoltage ratio defined as $(V_{AP} - V_P)/V_P$ is around 40%, which is similar with the TMR ratio of the MTJ.

shown in Fig. 2b a negative thermovoltage ΔV is detected in this geometry. While sweeping the in-plane applied field, a sudden ΔV jump is observed as the free layer magnetization switches and the magnetization configuration changes from P to AP. In fact the ΔV vs H hysteresis loop mimics the R vs H loop. Two thermovoltage levels are clearly defined corresponding to the two magnetization alignments (P and AP). The amplitude of the thermovoltage for the AP alignment, $\Delta V_{AP}$, is found to be larger than that for the P one, $\Delta V_P$. In our case, the $\Delta V_{AP}$ can reach up to − 1.07 mV while heating the top lead with a 125 mW laser power. The $\Delta V_{AP}$ is about 310 µV when heating the bottom lead with the same laser power (see Fig. 2b). This difference can be understood since different material, thickness and size for the top and bottom leads result in different heat conductivity and dissipation. In the case where the laser heats the bottom lead, i.e. $\Delta T < 0$, then a positive thermovoltage is measured as shown in Fig. 2b and an inverse thermovoltage ΔV hysteresis loop is observed. Note that if the laser is turned off or shines the substrate instead of the leads, the thermovoltage drops to zero and no influence of the applied field is observed.

For both top and bottom heating, the tunnel magnetothermovoltage ratio defined as $(\Delta V_{AP} - \Delta V_P)/\Delta V_P$ is around 40%, which is similar to the TMR ratio. This behavior suggests that the observed thermoelectric effect mainly results from the thermal spin-dependent tunneling between both sides of the $Al_2O_3$ barrier. Moreover the thermovoltage of the lead was measured as heating one end with the maximum laser power and a value lower than 2 µV was obtained. Thus, the thermovoltage of the lead can be neglected considering the measured thermovoltage in the MTJ.

Figure 3 shows the thermovoltage $\Delta V_P$ and $\Delta V_{AP}$ as a function of the laser power P in the cases of heating the top lead (Fig. 3a) and the bottom (Fig. 3b). One can see that the amplitudes of both $\Delta V_P$ and $\Delta V_{AP}$ increase with the laser power. The experimental data, obtained for $\Delta V_P$ and $\Delta V_{AP}$ are not linear but following a $P^{1/2}$ behavior for the top and bottom lead heating. In our understanding this $P^{1/2}$ behavior is not the signature of a microscopic process, but rather is due to the dependence of the temperature difference and of the Seebeck coefficient with P. Due to power dissipation, the temperature difference follows a $P^\alpha$ law with a lower than 1, and the Seebeck coefficient is itself a complicated function of the temperature, and therefore of $P$. The magnetothermovoltage ratio $(\Delta V_{AP} - \Delta V_P)/\Delta V_P$ is 40% which is close to the TMR ratio, and changes little with the laser power, as shown in Fig.3c.

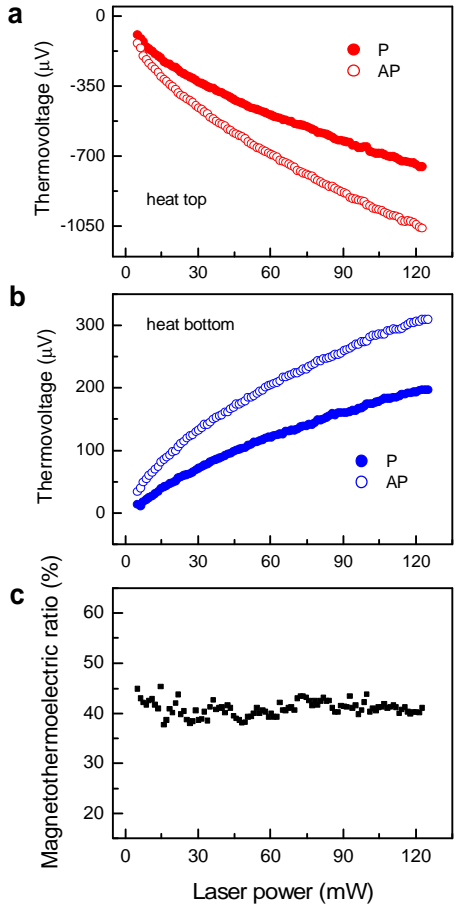

**Figure 3 Laser power dependences of the magnetic thermovoltage and magnetothermoelectric ratio in the MTJ.** The thermovoltage $\Delta V_P$ and $\Delta V_{AP}$ as a function of the laser power P in the cases of heating the top lead (**a**) and the bottom lead (**b**). It is found that $\Delta V_P$ and $\Delta V_{AP}$ are proportional to $P^{1/2}$. **c,** the magnetothermovoltage ratio $(\Delta V_{AP} - \Delta V_P)/\Delta V_P$ as a function of the laser power. It is around 40% which is very close to the TMR ratio, and is constant with the laser power.

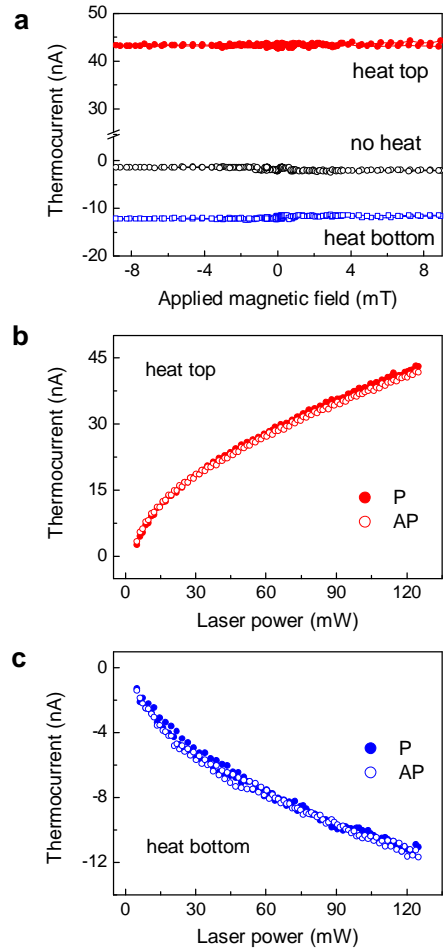

**Figure 4 Magnetic field and laser power dependence of the thermotunnel current. a,** Thermotunnel current vs. applied field (I-H) curves. I is independent on the magnetization alignments. **b, c,** The thermotunnel current $I_P$ and $I_{AP}$ as a function of the laser power P for both top and bottom lead heating It is found that $I_P$ and $I_{AP}$ are following a $P^{1/2}$ behavior.

Figure 4a shows the measured thermotunnel current $I$ as a function of $H$ in a closed-circuit geometry as described above. Without laser heating, the closed-circuit current is around zero. As heating the top lead with a 125 mW laser power, $I$ reaches 43 nA, and $I$ is about −12 nA when the bottom lead is heated by the same laser power. One can see that the thermotunnel current $I$ is independent on the magnetization alignments. However similarly with the thermovoltage, the laser power dependence of the thermotunnel current is also proportional to $P^{1/2}$ for both top and bottom leads heating, as shown in Fig. 4b and Fig. 4c. Considering the amplitude, the sign and the magnetic dependence of the measured signal, the possibility of an artifact coming from the known light-induced phenemona could be ruled out.

From the above experimental results obtained in $Al_2O_3$ MTJs one can see that the magnetothermovoltage is proportional to TMR, i.e. $\Delta V_P/\Delta V_{AP} = R_P/R_{AP}$, whereas the thermotunnel current is independent on magnetizations relative orientations. Since the thermovoltage is given by $\Delta V = -(G_T/G_V)\Delta T = -S\Delta T$ whereas the thermocurrent is given by $I = G_T \Delta T$, assuming that for a fixed laser power $\Delta T$ is constant for the P and AP configuration, we could define a tunnel thermopower $S$ which depends strongly on the magnetization alignment of the two magnetic layers. It leads to the conclusion that the coefficient $G_T$ is independent on the magnetization alignments. Consequently the tunnel magnetothermopower is proportional to TMR in $Al_2O_3$ MTJs, i.e. $S_P/S_{AP} = R_P/R_{AP}$. It should be said that this behavior is not conventional, since usually the Seebeck coefficient is not dominated only by the resistance. Indeed, the experiments of other groups[27,28] show that there is no direct relation between magneto-Seebeck effect and TMR in MgO MTJs, which agree with the ab initio calculation[26].

To obtain the value of $S$, the temperature difference between both sides of the tunnel barrier is needed. Unfortunately it is hard to directly measure a small temperature difference between a 2 nm barrier. Measuring the temperature of the leads using either with a thermocouple or the thermal variation of the leads resistance, we could conclude that the temperature difference is smaller than 1 K between the top and the bottom leads which could be much smaller between both sides of the tunnel barrier in the MTJ (see supplementary information). The tunnel thermopower $S$ in $Al_2O_3$ MTJ can therefore be estimated. A low estimate is 1 mV K$^{-1}$, which is large compared to conventional metals and semiconductors[32]. Walter et al.[27] show the thermopower in MgO MTJ of 100 (1300) μV K$^{-1}$ with a 5.3 μV measured thermovolatge and a simulated temperature difference across the barrier of 53 mK (4.4 mK). Theoretical study by McCann et al.[25] using inelastic magnon model estimates 55 μV K$^{-1}$ and the ab-initio theory by Czerner et al.[26] gives values of up to 150 μV K$^{-1}$.

This means that giant thermopower can be obtained in an $Al_2O_3$ MTJ.

In the following we provide a simple model within the linear response theory that agrees with our experimental results. Note that this explanation does not exclude that this behavior could result from a very peculiar inelastic scattering of the electrons with phonons or magnons, but in the absence of detailed experimental evidences of this type of process we will use a description based on elastic scattering only and find the particularities needed to explain the experimental data. In such a case it is possible to express the Onsager coefficient $L_{11} = G_V$ and $L_{12} = G_T$ as the moments of order 0 and 1 of the transport function $s(e)$,

$$L_{11} = \int de\, s(e)\left(-\frac{\partial f}{\partial e}\right) \quad L_{12} = \frac{1}{(-e)T}\int de\, s(e)(e-m)\left(-\frac{\partial f}{\partial e}\right) \quad (2)$$

$f$ is the Fermi-Dirac distribution function, and (-e) the electron charge. This approach is well used for bulk – thermoelectricity[22,32,33] and has recently been applied in the context of spin caloritronics in the works by Czerner et al[26] and Walter et al[27]. The function $s(e)$ has the physical meaning of an energy-dependent conductivity for the electrons. The quantities $L_{11}$ and $L_{12}$ measure respectively the value and the slope of the function $s(e)$, $k_B T$ around the Fermi level.

In the case of a P configuration, $s(e)$ is given by

$$s_P(e) = \frac{2p}{h}e^2\left\{|T_{\uparrow\uparrow}|^2 r_\uparrow^L r_\uparrow^R + |T_{\downarrow\downarrow}|^2 r_\downarrow^L r_\downarrow^R\right\} \quad (3)$$

whereas for an AP configuration,

$$s_{AP}(e) = \frac{2p}{h}e^2\left\{|T_{\uparrow\downarrow}|^2 r_\uparrow^L r_\downarrow^R + |T_{\downarrow\uparrow}|^2 r_\downarrow^L r_\uparrow^R\right\} \quad (4)$$

$r_\uparrow^{L,R}$ and $r_\downarrow^{L,R}$ are the spin-up and spin-down density of states (DOS) in the left (L) and right (R) leads. $|T_{ss}|^2$ are the tunneling functions. From $S = G_T/G_V = L_{12}/L_{11}$ it is clear that the thermopower will be proportional to the resistivity $1/L_{11}$, if $L_{12}$ is independent of the magnetization orientation, as found in our experiment. In view of equation (2) this requires the slope of the transport function, averaged $k_B T$ around the Fermi level, to be the same in the P and AP configuration. Notice that this does not preclude at all for the values of $s_P$ and $s_{AP}$ to be different and therefore allow observing a TMR.

Unlike for the MgO MTJs, the Julliere model[34] may be appropriate for the $Al_2O_3$ MTJs. Therefore neglecting the energy dependence of the tunneling functions, the slopes of $r_\uparrow^L r_\uparrow^R + r_\downarrow^L r_\downarrow^R$ and $r_\uparrow^L r_\downarrow^R + r_\downarrow^L r_\uparrow^R$ should then approximately be the same. Our experimental results would be consistent with DOSs written as $r_\uparrow = r_{0\uparrow} + dr$ and $r_\downarrow = r_{0\downarrow} + dr$ where $r_{0\uparrow}$ and $r_{0\downarrow}$ are the DOS for an alloy of cobalt with iron, and $dr$ a spin independent contribution which can be understood as a resonance. In such a case, with $r_0^{L,R} = r_{0\uparrow}^{L,R} + r_{0\downarrow}^{L,R}$

$$s_P \propto r_{0\uparrow}^L r_{0\uparrow}^R + r_{0\downarrow}^L r_{0\downarrow}^R + dr(r_0^L + r_0^R) + 2dr^2 \quad (5)$$

$$s_{AP} \propto r_{0\uparrow}^L r_{0\downarrow}^R + r_{0\downarrow}^L r_{0\uparrow}^R + dr(r_0^L + r_0^R) + 2dr^2 \quad (6)$$

Because spin-up and spin-down DOS of bulk cobalt and iron have small slope at the Fermi level on the scale of $k_B T$, it is a good approximation that it is also true for their alloys, if no special atomic order is created, as in our compounds. The energy dependence and the slopes of $S_P$ and $S_{AP}$ are then dominated by the one of the resonance $dr$, and therefore is independent on the P or AP configuration. Inserting equations (5) and (6) into equations (2), we find the Seebeck coefficients

$$S_P = \frac{p^2 k_B^2}{3(-e)} T \frac{dr'/r^L + dr'/r^R}{a^R a^L + (1-a^R)(1-a^L)} \quad (7)$$

$$S_{AP} = \frac{p^2 k_B^2}{3(-e)} T \frac{dr'/r^L + dr'/r^R}{(1-a^R)a^L + a^R(1-a^L)} \quad (8)$$

To obtain these expressions we have used the low temperature expansion of the Fermi function, and the following definitions $r^{L,R} = r^{L,R}_\uparrow + r^{L,R}_\downarrow$, $a^{L,R} = r^{L,R}_\uparrow / r^{L,R}$.

The numerator of equations (7) and (8) describes why the thermocurrent of Fig. 4a is independent of the magnetization alignment whereas in the denominator we recognize the Jullière expression for the conductance in term of polarizations $a^{L,R}$. This explains the proportionality observed between the magneto-Seebeck effect and TMR in Fig. 2. These formula are also consistent with the large value observed for the thermovoltage if $dr$ is a very narrow nonmagnetic resonance giving rise to a large $dr'$. This could originate for example from nonmagnetic impurity states, since they are usually narrow.

In summary, large thermoelectric effect was observed in the MTJ arising from the temperature difference between both sides of a 2 nm $Al_2O_3$ tunnel barrier. The magnetothermovoltage ratio for the P and AP magnetization configuration is similar to the TMR ratio in the $Al_2O_3$ MTJ. However the thermotunnel current is independent on the magnetization alignments. The thermopower can be estimated to be larger than 1 mV K$^{-1}$ in the $Al_2O_3$ MTJ which is larger than that in the metal and semiconductor, suggesting that MTJ can be used as a good thermospin device. The thermospin devices can work in an open-circuit without applying any current or voltage. On one hand, the large change in thermovoltage can be obtained in the presence of a temperature difference through controlling the relative magnetization alignment of the two ferromagnetic layers in the MTJ. On the other hand, the magnetothermovoltage can be used to detect the magnetization configuration even in the open-circuit geometry. The exact mechanism may still be discussed but we are proposing a description based on elastic scattering to explain qualitatively the experimental results.

This work extends the understanding of the spin-dependent thermal and electrical transport in nanostructures, and has promising potential for the design and application of thermally driven magnetic tunnel junction.

## Methods

### MTJ preparation
The MTJ consists of a bottom reference layer Ta(5 nm)/PtMn(25 nm)/Co$_{90}$Fe$_{10}$(2 nm)/Ru(0.8 nm)/Co$_{90}$Fe$_{10}$(3 nm) and a free layer Co$_{90}$Fe$_{10}$(2 nm)/Ni$_{80}$Fe$_{20}$(5 nm)/Ru(4.8 nm)/Au(10 nm) separated by a 2 nm thick $Al_2O_3$ barrier. The films were deposited on the 400 nm $Al_2O_3$ covered Si wafers in a dc magnetron sputtering system at RT with a base pressure of $2\times10^{-8}$ Torr and a deposition pressures of 2–3 mTorr. The $Al_2O_3$ barrier was obtained by reactive rf oxidation of a 2 nm Al layer at a power of 50 W. The films were annealed for 2 hours at the temperature of 265 °C and a 1.3 T magnetic field in a $N_2$ atmosphere oven, and then patterned to circular shape with the diameter varying from 40 to 100 μm using the photolithography and ion mill processes. The 200 nm Cu and 10 nm Ta were used as both the bottom and top leads. The MTJs were measured using a system sourcemeter. The TMR is around (40 ± 3)% and the resistance-area (RA) product is about (22 ± 6) MΩμm$^2$ at RT.

### Magnetothermovoltage and thermotunnel current measurements
To generate a temperature difference between the reference layer and the free layer, the top lead or the bottom was heated using a laser beam from a laser diode with a wavelength of 780 nm and a maximum power of 125 mW. The laser spot on the lead is around 5 mm away from the junction. It should be noted that only part of heat pass through the MTJ, because the power is dissipated and the size of the MTJ is much smaller than that of the leads. The thermovoltage was measured by a nanovoltmeter having an internal resistance larger than 10 GΩ in an open-circuit at RT with an applied magnetic field up to 0.3 T along the in-plane easy axis. The thermotunnel current was measured by a sourcemeter having an internal resistance lower than 100 mΩ connecting the 16 kΩ MTJ in a closed-circuit without applied voltage.

The thermovolage of MTJ was also checked by measuring the AC voltage using a lock-in with the same frequency of the AC laser power. The AC measurement shows the similar behavior with the DC measurement.

We measured the temperature difference in between the bottom and top leads using k-type thermocouples connected with the nanovoltmeter. For a laser power of 125 mW, the temperature of the leads are about 319 K, and the temperature difference between the top lead and the bottom one is 300±250 mK (see Supplementary Information). Because the size of MTJ is much smaller than that of the leads and the 2 nm thinckness of $Al_2O_3$ barrier is small comparing to the 60 nm thickness of the whole multilayers. the temperature difference across the barrier could be much smaller and estimated to be in the order of 100 mK.

The $Al_2O_3$ MTJs with the diameters varying from 40 to 100 μm were measured and showed similar behavior. (See Supplementary Information)

### Acknowledgements
The authors thank French National Research Agency (ANR) PNANO and ISTRADE for support. W. L. and S. M. thank J. Z. Sun, Eric E Fullerton and A.D. Kent for the fruitful discussion.

### Author contributions
W.L. conceived and designed the experiments, and performed the measurements and analysis. S.M. supervised the experiment. B.N. prepared the MTJ samples in this paper. M.H. and S.A. prepared the additional MTJ films and F.M. patterned the additional MTJ samples. W.L. and S.M. gave physical explanations and L.C. contributed for the theoretical model. M.H. and S.A. contributed to the helpful discussions. W.L., L.C. and S.M. wrote the paper.

# Supplementary Information

## Giant thermoelectric effect in Al$_2$O$_3$ magnetic tunnel junctions

Weiwei Lin, Michel Hehn, Laurent Chaput, Béatrice Negulescu, Stéphane Andrieu, François Montaigne and Stéphane Mangin

Institut Jean Lamour, Nancy-Université, Vandoeuvre-lès-Nancy 54506, France

### Magnetothermoelectric effect in Co$_{90}$Fe$_{10}$/Al$_2$O$_3$/Co$_{90}$Fe$_{10}$ tunnel junctions

We studied the thermoelectric effect for tens of Al$_2$O$_3$ based magnetic tunnel junctions (MTJs). Several devices with the same stack but various sizes were measured.

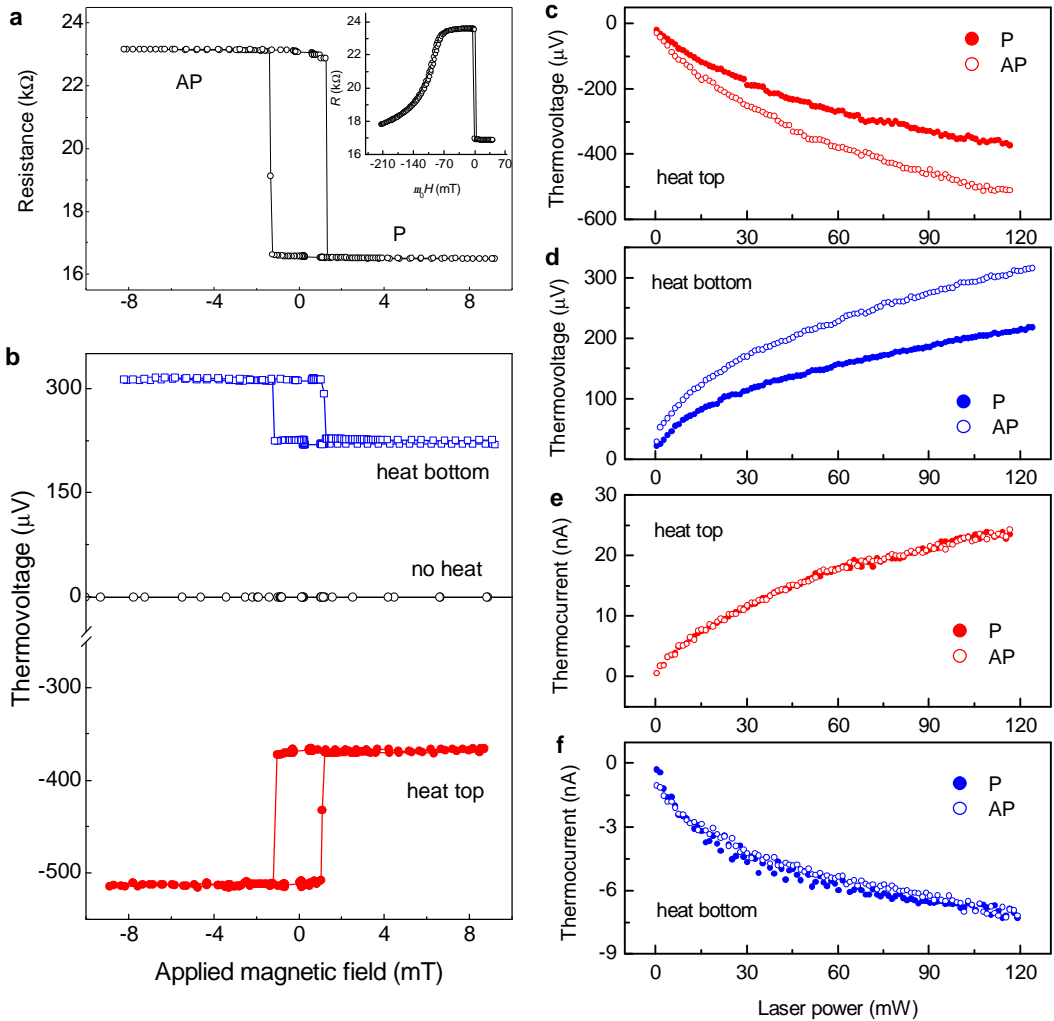

**Figure S1 a,** Magnetic minor loop of the tunnel resistance $R$ in a Ta(5 nm)/PtMn(25 nm)/Co$_{90}$Fe$_{10}$(2 nm)/Ru(0.8 nm)/Co$_{90}$Fe$_{10}$(3 nm)/Al$_2$O$_3$(2 nm)/Co$_{90}$Fe$_{10}$(2 nm)/Ni$_{80}$Fe$_{20}$(5 nm)/Ru(4.8 nm)/Au(10 nm) MTJ with the diameter of 80 µm. The inset is the magnetic major loop of $R$. **b,** Magnetic loops of the thermovoltage $\Delta V$ in the MTJ. The red curve shows the case that the top lead is heated by the 125 mW laser, and the blue one indicates the case that the bottom lead is heated with the same laser power. **c,d,** Laser power $P$ dependences of the thermovoltages $\Delta V_\text{P}$ and $\Delta V_\text{AP}$ in the cases of heating the top lead and the bottom, respectively. **e,f,** Laser power dependences of the thermotunnel currents $I_\text{P}$ and $I_\text{AP}$ in the cases of heating the top and the bottom lead, respectively.

Figure S1 shows the minor loops of the tunnel resistance and the thermovoltage as a function of the applied magnetic field in a Ta(5 nm)/PtMn(25 nm)/Co$_{90}$Fe$_{10}$(2 nm)/Ru(0.8 nm)/Co$_{90}$Fe$_{10}$(3 nm)/Al$_2$O$_3$(2 nm)/Co$_{90}$Fe$_{10}$(2

nm)/Ni$_{80}$Fe$_{20}$(5 nm)/Ru(4.8 nm)/Au(10 nm) MTJ with the diameter of 80 μm, which has the same stack and same size with the one shown in the paper. It gives very similar results: The tunnel resistance for the parallel (P) magnetization configuration $R_P$ is 16.5 kΩ, and the TMR ratio is 40.6%. As the top lead is heated by the 125 mW laser, the open-circuit thermovoltage for the antiparallel (AP) magnetization configuration $\Delta V_{AP}$ is –514 μV, and the tunnel magnetothermovoltage ratio is 40.2%. The $\Delta V_{AP}$ is 315 μV as heating the bottom lead by the 125 mW laser, and the magnetothermovoltage ratio is 40.0%. $\Delta V_P$ and $\Delta V_{AP}$ is following a $P^{1/2}$ behaviour while heating the top lead and the bottom. The tunnel thermocurrent measured in a closed-circuit is 23 nA as heating the top lead with 125 mW laser power and is –7 nA as heating the bottom lead, with no visible difference for the P and AP configurations.

Note that for different pieces of MTJs, the thermovoltage can be different while heating the sample with the same laser power. This variation can be understood since the location of the laser spot is not well controlled and heat dissipation may be different from one sample to another and since the thermovoltage is strongly related to the temperature difference between both sides of the 2 nm tunnel barrier.

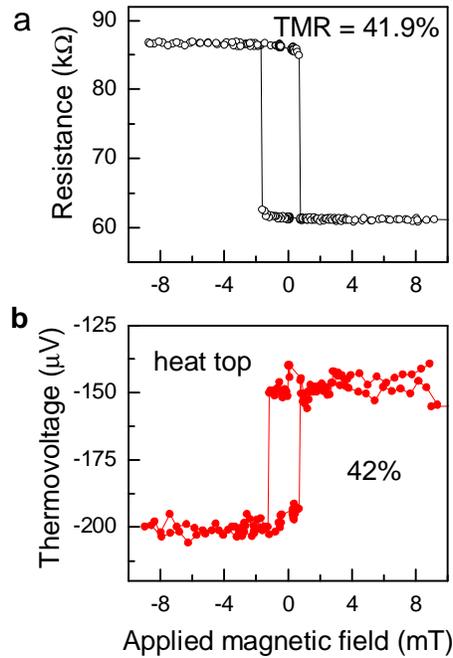

**Figure S2 a,** Magnetic minor loop of the tunnel resistance in a Ta(5 nm)/PtMn(25 nm)/Co$_{90}$Fe$_{10}$(2 nm)/Ru(0.8 nm)/Co$_{90}$Fe$_{10}$(3 nm)/Al$_2$O$_3$(2 nm)/Co$_{90}$Fe$_{10}$(2 nm)/Ni$_{80}$Fe$_{20}$(5 nm)/Ru(4.8 nm)/Au(10 nm) MTJ with the diameter of 40 μm. **b,** Magnetic loop of the thermovoltage as heating the top lead with the 125 mW laser.

The Al$_2$O$_3$ MTJs with diameters varying from 40 to 100 μm were measured. Figure S2 shows the magnetic minor loops of the tunnel resistance and the thermovoltage in a Al$_2$O$_3$ MTJ with a diameter of 40 μm. Similar behaviours are observed. The resistance $R_P$ is 61.1 kΩ, and the TMR ratio is 41.9%. The thermovoltage $\Delta V_{AP}$ is –203 μV as the top lead is heated by the 125 mW laser, and the magnetothermovoltage ratio is about 42% which is again very close to the TMR ratio.

These results confirm that the large amplitude of thermovoltage can be observed for different Al$_2$O$_3$ MTJs size, that the magnetothermovoltage ratio is very close to the TMR ratio and that thermocurrent is constant for the P and AP configurations.

**Laser power dependence of the temperature difference of the leads in $Co_{90}Fe_{10}/Al_2O_3/Co_{90}Fe_{10}$ tunnel junctions**

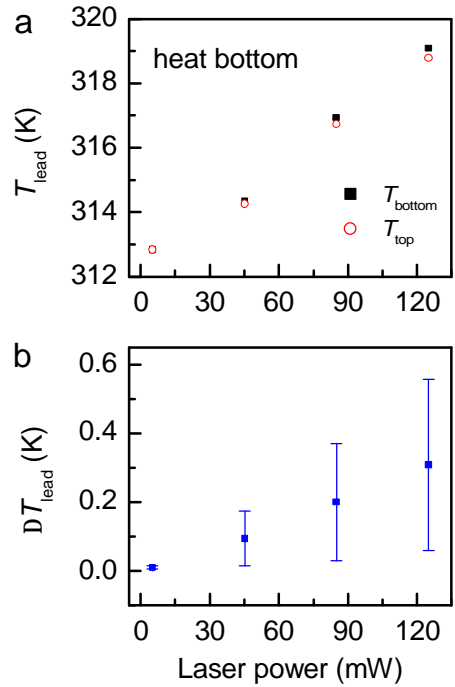

**Figure S3 a,** Temperature of the bottom and top leads as the bottom lead was heated for various laser powers in a $Al_2O_3$ MTJ. **b,** Laser power dependence of the temperature difference between the bottom and the top leads.

The temperature difference between the bottom and top leads was obtained by measuring the temperature of each lead with a k-type thermocouple connected to a nanovoltmeter. To minimise the error, we tried to install the thermocoupled as close as possible to the junction. In order to estimate the error, we repeated the measurments which allowed us to give error bar on the temperature difference in Fig. S3. For instance with a laser power of 125 mW heating the bottom lead, the temperatures of the leads are about 319 K, whereas the temperature difference between the top lead and the bottom one is 300±250 mK. It should be noted that only part of the heat goes through the MTJ, and that the vertical temperature difference across the 2 nm $Al_2O_3$ barrier should be smaller than the one measured between the two leads. For a 125 mW laser power, an upper limit for the temperature difference across the barrier can be estimated to be about 100 mK. Consequently from the measured 1 mV thermovotage, we can estimate that the Seebeck coefficients in $Al_2O_3$ MTJ should be on the order of 1 mV $K^{-1}$ or larger.